\documentclass[prl,twocolumn]{revtex4}
\usepackage{bm}
\usepackage{graphicx}
\usepackage{amsmath}
\usepackage{eufrak}
\usepackage{color}
\usepackage{ulem}
\newcommand{\nix}[1]{}

\begin{document}

\title{Fast detector of  the ellipticity of infrared and terahertz 
radiation based on HgTe quantum well structures}

\author{S.\,N.\,Danilov,$^1$ B.\,Wittmann,$^1$ P.\,Olbrich,$^1$   
W.\,Eder,$^1$ W.\,Prettl,$^1$ L.\,E.\,Golub,$^2$ E.\,V.\,Beregulin,$^2$
Z.\,D.\,Kvon,$^3$ N.\,N.\,Mikhailov,$^3$ S.\,A.\,Dvoretsky,$^3$ 
V.A.\,Shalygin,$^4$  N.\,Q.\,Vinh,$^5$ A.\,F.\,G.~van~der~Meer,$^5$ B.~Murdin,$^6$ 
and S.\,D.\,Ganichev$^{1}\footnote{e-mail: sergey.ganichev@physik.uni-regensburg.de}$}
\affiliation{$^1$  Terahertz Center, University of Regensburg, 93040
Regensburg, Germany}
\affiliation{$^2$A.F.~Ioffe Physico-Technical Institute, Russian
Academy of Sciences, 194021 St.~Petersburg, Russia}
\affiliation{$^3$ Institute of Semiconductor Physics, Russian Academy 
of Sciences, 630090 Novosibirsk, Russia}

\affiliation{$^4$ ~St. Petersburg State Polytechnic University, 195251
St.~Petersburg, Russia}

\affiliation{$^5$ FOM Institute for Plasma Physics ``Rijnhuizen'', P.O. 
Box 1207, NL-3430 BE Nieuwegein, The Netherlands}

\affiliation{$^6$University of Surrey, Guildford, GU2 7XH, UK}

\pacs{42.25.Ja,29.40.-n,78.67.-n,78.67.De}



\begin{abstract}
We report a fast, room temperature detection scheme for 
the polarization ellipticity of laser radiation, with a
 bandwidth that stretches from the infrared to the terahertz range.  
 The device consists of two elements, one in front of the other, that 
 detect the polarization ellipticity and the azimuthal angle of the 
 ellipse.  The elements respectively utilise the circular photogalvanic 
 effect in a narrow gap semiconductor and the linear photogalvanic effect 
 in a bulk piezoelectric semiconductor.  For the former we characterized 
 both a HgTe quantum well and bulk Te, and for the latter, bulk GaAs.  
 In contrast with optical methods our device is an easy to handle 
 all-electric approach, which we demonstrated by applying a large number 
 of different lasers from low power, continuous wave systems to high 
 power, pulsed sources.

\end{abstract}

\maketitle

\section{Introduction}

Fast and easy recording of the state of polarization, i.e. measurements of the Stokes parameters 
of a radiation field is of great importance for various applications in science and 
technology. In particular the ellipticity  of transmitted, reflected or scattered light
may be used to analyze the optical anisotropy of a wide range of media. 
The established method for gaining information about the polarization state 
of light is the use of optical elements to determine 
optical path differences. 
Recently  we presented a new technique which allows  
\textit{all-electric} room temperature detection of the state
of polarization providing  full characterization of laser beams at 
THz frequencies~\cite{APL2007,JAP08}. The operation of the detector system is
based on photogalvanic effects in  semiconductor quantum well
(QW) structures of suitably low symmetry~\cite{bookIvchenko,book}. 
The detection principle has been demonstrated on doped GaAs and 
SiGe QWs at room temperature applying terahertz molecular lasers as 
radiation sources~\cite{APL2007,JAP08}. The time constant of photogalvanic currents 
is determined by the momentum relaxation time of free carriers which is in the range of
picoseconds at room temperature. This makes possible to measure  the
state of polarization of laser radiation with sub-nanosecond time resolution. 
Here we report a substantial improvement of the sensitivity of the method, 
by about two orders of magnitude higher compare to those previously reported in GaAs QWs, 
and on the extension of the detector's spectral range from teraherz to the mid-infrared in a single device.
To achieve these goals we studied photogalvanic effects 
in QWs prepared from the narrow gap semiconductor HgTe and also analyzed 
bulk single-crystalline Te and GaAs.
HgTe QWs are  novel promising narrow gap materials characterized by   high electron mobilities, 
low effective masses, an inverted band structure,  large
$g$-factors, and spin-orbit splittings of subbands in the momentum space~\cite{Pfeuffer,Zhang,Zhang2,Gui,Kvon}. 
Because of these features low dimensional HgTe/CdHgTe structures hold a
great potential  for detection of terahertz radiation~\cite{book,sakai05,miles07,woolard07,edwards00,shur03}
and for the rapidly developing field of spintronics~\cite{Ch7spintronicbook02,Zutic04review,Dyakonov08}.
The most important property of the materials relevant for the detection of the radiation ellipticity 
is the magnitude of the  circular photogalvanic effect (CPGE)~\cite{APL2007,JAP08,book}
which gives access to the helicity of a radiation field. Therefore we focus our work on 
the study of the CPGE. We show that  HgTe QWs can be used for all-electric detection of
radiation ellipticity in a wide spectral range from far-infrared (terahertz radiation) 
to mid-infrared wavelengths. The detection is demonstrated for various low 
power $cw$ and high power $pulsed$ laser systems.
A large spectral range was covered by 
low-pressure cw, pulsed, and Q-switched CO$_2$ lasers and high-power 
pulsed transverse excited atmospheric pressure (TEA) CO$_2$ lasers, which are used directly 
in the infrared range or as pump sources for THz molecular lasers. 
In addition measurements were carried out 
with  a free-electron-laser making use of its tunability and short pulses.

\section{Experimental}

\begin{figure}[h]
\includegraphics[width=0.9\linewidth]{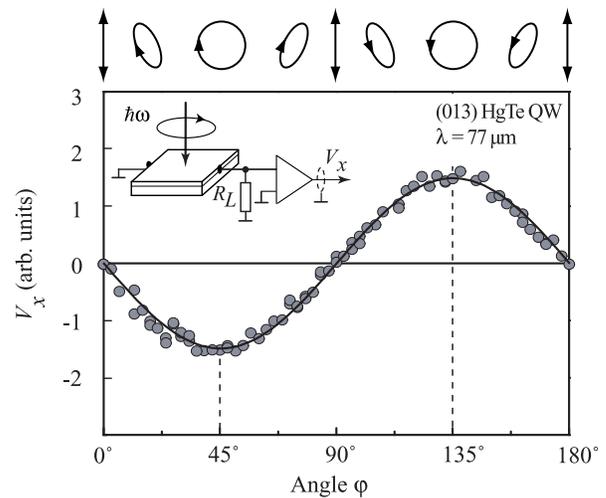}
\caption{ 
Helicity dependence of the photoresponse normalized by the radiation power $V_x/P$ in a (013)-grown HgTe 
QW at room temperature. The signals are obtained at $\lambda = 77~\mu$m applying 
pulsed radiation of NH$_3$ THz laser.
The full line is a fit after Eq.~\protect\eqref{CPGE1}. 
The inset shows the experimental geometry. On top of the figure polarization ellipses  corresponding to various phase
angles $\varphi$ are plotted viewing from the direction toward which the wave approaching.
}
 \label{fig1}
\end{figure}

The experiments were carried out on Cd$_{0.7}$Hg$_{0.3}$Te/HgTe/Cd$_{0.7}$Hg$_{0.3}$Te 
single QWs of 21~nm width.
The structures were grown on GaAs substrates with  surface orientation (013), by 
means of a modified Molecular Beam Epitaxy method~\cite{MBE}.  
Samples with   electron densities   of
$9 \times 10^{11}$~cm$^{-2}$ 
and mobility   
$2.5 \times 10^4$~cm$^2$/Vs 
are studied  at room temperature. 
We used square shaped samples of 5$\times$5~mm$^{2}$ size.
Two pairs of contacts (along directions $x$ and $y$) was centered in the 
middle of cleaved edges parallel to the intersection of the (013) plane 
and cleaved edge face \{110\} (see the inset to Fig.~\ref{fig1}). 
To demonstrate the detector operation in the
mid-infrared spectral range we used a Q-switched CO$_2$ laser covering the spectrum
from 9.2 to 10.8~$\mu$m with peak power $P$ of about 1~kW and pulsed  TEA-CO$_2$ laser generating single 100~ns pulses 
of power up to 100~kW~\cite{book}. Besides gas lasers 
we  used the output from the free electron laser ``FELIX'' at 
FOM-Rijnhuizen in the Netherlands   
at wavelengths  between 5~$\mu$m and 17~$\mu$m and power about 
100~kW~\cite{Knippels99p1578}. Making use of the
frequency tunability and short pulse duration of this laser we obtain the spectral 
behaviour of the detector responsivity and demonstrated
its time resolution. The output pulses of light
from FELIX were chosen to be $\approx$~3~ps long, separated by 40~ns, in a
train (or ''macropulse'') of duration of 7~$\mu$s. The macropulses
had a repetition rate of 5~Hz. For optical excitation in the terahertz range we used 
a $cw$  optically pumped CH$_3$OH laser with  
radiation wavelength of 118~$\mu$m 
and power $P$ of about 10~mW and  high power  
pulsed optically pumped NH$_3$, D$_2$O, and CH$_3$F  lasers~\cite{book}.
Several lines of the pulsed laser with power  ranging from 100~W 
to 100~kW in the wavelength range between 
$\lambda = 76$~$\mu$m and 496~$\mu$m have been applied. 

To vary the ellipticity of the laser beam  we used  $\lambda/4$ plates  
made of   $x$-cut crystalline quartz in the terahertz range 
and a Fresnel rhomb in the mid-infrared.
The light  polarization  was varied from linear to elliptical.
Rotating the polarizer varies the
helicity of the  light, $P_{\rm circ} =  \sin{2 \varphi}$,   from
$P_{\rm circ} =-1$ (left-handed circular, $\sigma_-$) to $P_{\rm circ}
=+1$ (right-handed circular, $\sigma_+)$ where
$\varphi$ is the angle between the initial linear polarization and
the optical axis (c-axis) of the polarizer.
In  Fig.~\ref{fig1}~(top)  the shape of the polarization ellipse and the handedness of 
the radiation
are shown for various angles $\varphi$.

Photogalvanic effects applied here for radiation detection generate an electric current which we measure in
homogeneously irradiated  unbiased 
samples.  The current is converted into a signal voltage by load resistors.
The signal voltages were picked
up from the detector 
across a $R_L$~=~50~$\Omega$ load resistor for short radiation pulses. For cw irradiation  we use 
a load resistance of  1~M$\Omega$, which is much higher than the sample resistance. 
For pulsed lasers the signals were fed into amplifiers with voltage amplification by 
a factor of 100  and a bandwidth of
300~MHz and were recorded  by a digital  broad-band (1~GHz) oscilloscope. 
For time resolved measurements the signal was picked up directly from the sample 
without amplification. For detection of
$cw$ laser radiation we modulated our beam with a chopper at modulation frequency 
225~Hz and used a low-noise pre-amplifier
(100 times voltage amplification) and a lock-in-amplifier for signal recording. 
We note that in this low modulation frequency
case the high time resolution of the set-up is not required.

\begin{figure}[h]
\includegraphics[width=0.9\linewidth]{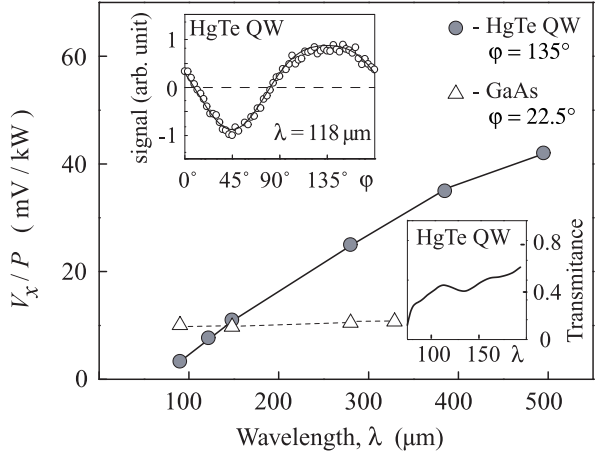}
\caption{CPGE spectrum of HgTe QW (full circles) in the THz-range normalized by power 
incident on the sample for circularly polarized  radiation ($\varphi = 135^\circ$). 
Top inset shows the helicity dependence obtained 
in HgTe QW applying radiation of a 
$cw$ CH$_3$OH laser at $\lambda = 118$~$\mu$m.  
Full line in the inset is the fit after Eq.~(\protect \ref{hgte1}).
Bottom inset shows transmittance spectrum obtained 
by FTIR spectroscopy. Triangles show the spectrum of bulk GaAs sample in 
response to the elliptically polarized radiation. 
The data are given for angles $\varphi = 22.5^\circ$ corresponding to the
maximum of the voltage signal.
}
 \label{fig3}
\end{figure}%

\begin{figure}[h]
\includegraphics[width=0.9\linewidth]{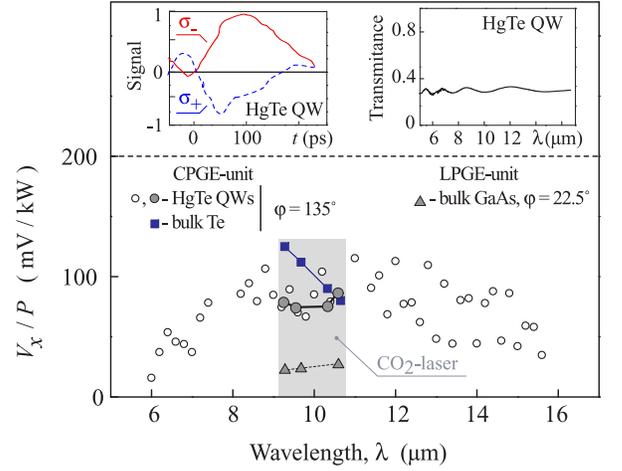}
\caption{ CPGE spectrum of HgTe QW and  Te bulk crystal obtained in response to the  
the mid-infrared circular polarized radiation
and normalized by power incident on the sample for circularly polarized  radiation. 
The data are obtained for angles $\varphi = 135^\circ$ 
applying Q-switched CO$_2$ laser 
(full circles) and free electron laser FELIX (open circles). The data of Te are plotted 
by squares. Triangles represent spectral 
response of the LPGE bulk GaAs unit obtained by pulsed TEA-CO$_2$ laser radiation. 
The data are given for angles $\varphi = 22.5^\circ$ corresponding to the
maximum of the voltage signal. The shadow area indicates the spectral range of a CO$_2$ laser.
Left and right insets show the CPGE temporal response of HgTe 
QWs to sub-ps infrared pulses of FELIX ($\lambda$~=~9~$\mu$m) and 
the transmittance spectra obtained by FTIR spectroscopy, respectively.
}
 \label{fig2}
\end{figure}%

\section{Detection of the helicity applying H\lowercase{g}T\lowercase{e} QW\lowercase{s}}

With illumination of HgTe samples at normal incidence we detected in the in-plane $x$-direction
a helicity dependent current signal. This is shown in Fig.~\ref{fig1} for a measurement at room temperature 
obtained at 77~$\mu$m wavelength of a pulsed NH$_3$ laser. 
Note that the measurement is d.c. coupled and no background subtraction has occurred. 
The signal changes direction if 
the circular polarization is switched from left to right-handed 
($\sigma_-$ to $\sigma_+$ as shown in Fig.~\ref{fig1})
and vice versa.  
The helicity dependence and in particular the change of sign demonstrate that the observed 
current $j_x$ is due to the CPGE~\cite{book}. The voltage signal resulting from the 
photocurrent $V_x \propto j_x$ is well described by 
\begin{equation}
\label{CPGE1} 
V_x^{\rm HgTe} = S(\omega) P \cdot P_{\rm circ},
\end{equation}
where $S(\omega)$ denotes the strength of CPGE at radiation frequency $\omega$, and, 
consequently, the sensitivity of this detector unit.   
The photogalvanic current $\bbox{j}$ is  described by 
the phenomenological theory. It  can be written as a function of the 
electric component of the radiation field $\bm{E}$ and the propagation direction $\hat{\bm{e}}$ 
in the following form~\cite{bookIvchenko,Dyakonov08}
\begin{equation}
\label{photogalv} 
j_{\lambda} =
\sum_{\rho}
\gamma_{\lambda \rho}\:\hat{e}_\rho P_{\rm circ}\:|E|^2 +
\sum_{\mu, \nu}
\chi_{\lambda \mu \nu} (E_{\mu} E^*_{\nu} + E_{\mu}^* E_{\nu})\:,
\end{equation}
where the first term on the right hand side being proportional to the helicity 
or circular polarization degree $P_{\rm circ}$ 
of the radiation  represents the CPGE while the second term corresponds to the linear 
photogalvanic effect (LPGE)~\cite{book} which may be superimposed to the CPGE. 
The indices $\lambda$, $\rho$, $\mu$, $\nu$ run over the coordinate axes $x$, $y$, $z$. The second 
rank pseudotensor $\gamma_{\lambda \rho}$ and the third rank tensor $\chi_{\lambda \mu \nu}$ symmetric 
in the last two indices are material parameters. The  (013)-oriented QWs belong to the trivial 
point group C$_1$  lacking any symmetry operation except the identity.
Hence symmetry does not impose any  restriction on the relation between irradiation and 
photocurrent. All components of the tensor $\chi_{\lambda \mu \nu}$ and the pseudotensor 
$\gamma_{\lambda\rho}$ may be different from zero. 
Equation~(\ref{photogalv}) yields for normal incidence of the radiation the angle
$\varphi$ dependence of the photosignal $V_x \propto j_x$
\begin{equation}
\label{hgte1}
V_x/P = S(\omega) \sin 2\varphi + b(\omega) \sin 4\varphi + c(\omega) \cos 4\varphi + d(\omega),          
\end{equation}
where $S(\omega)$, $b(\omega)$, $c(\omega)$, and $d(\omega)$ can be consistently expressed by components 
of the tensors $\gamma_{\lambda \rho}$ and $\chi_{\lambda\mu\nu}$  
defined in Eq.~(\ref{photogalv})~\cite{condmatHgTe}.  
While at most wavelengths we found that the first term in Eq.~\eqref{hgte1} dominates (see Fig.~\ref{fig1}), at some wavelengths
we also find a small polarization independent current offset 
and a slight distortion of the sine-shape of the $V_x$ versus $\varphi$ curve. 
This deviation from the $\sin 2\varphi$ behaviour is demonstrated in the 
inset of Fig.~\ref{fig3} for $118~\mu$m wavelength of the $cw$ CH$_3$OH laser. 
The solid line in this inset is calculated after  Eq.~(\ref{hgte1})
yielding a good agreement to the experiment. Note, however, that for a technical 
application as an ellipticity detector described here it is desirable to 
have samples with dominating CPGE term being simply proportional to $P_{\rm circ}$ 
like the photoresponse in Fig.~\ref{fig1}. For all wavelengths this goal could  be achieved by using 
technologically available (112)-grown samples~\cite{Zhang}. The reason is the higher symmetry 
of such structures. 
In (013)-oriented samples  investigated in our work the CPGE current can have  
arbitrary direction and, in fact,  the projection of the current on the connecting line 
between the contacts  is measured. Furthermore three LPGE current contributions are allowed due to the low symmetry of the samples.
In contrast, in 
(112)-grown structures the current direction for normal incidence is bound to the 
[1$\bar{1}$0]-crystallographic direction. This is because  
such samples belong to the C$_s$ point group forcing 
the CPGE current to a direction normal to the mirror reflection plane. 
This feature makes the  preparation of samples with dominating CPGE
for all wavelengths possible.
Picking up the signal from [1$\bar{1}$0]-direction 
should increase the sensitivity 
and smoothen the  spectral dependence.

In Fig.~\ref{fig3} and Fig.~\ref{fig2} the spectral dependence of the CPGE is shown for the 
infrared and THz ranges, respectively. The measurements are carried out  
at room temperature at normal incidence of radiation and for  circularly polarized radiation 
($\varphi=135^\circ$) where the CPGE signal is  maximal. In the THz range 
molecular lasers were applied while the measurements in the infrared range
were performed using FELIX (open circles in Fig.~\ref{fig2}) and a Q-switched CO$_2$ laser 
(full circles in Fig.~\ref{fig2}). 
In the infrared range  CPGE in HgTe QWs does not show much dispersion for $\lambda$ between 
8 and 14~$\mu$m while in the THz range 
the photocurrent rises significantly with increasing wavelength. The spectral behaviour of the signal
is mainly caused by the mechanism involved in the radiation absorption, interband in the mid-infrared range 
and Drude-like in the terahertz range. 
The sensitivity in the mid-infrared range being of the order of 80~mV/kW 
is comparable with that of  photon drag detectors widely applied  
for detection of infrared laser radiation~\cite{book,Ryvkin80,Gibson80}.
Comparing  sensitivities of HgTe QW and GaAs QW devices in the THz range we obtain that 
a detector element made of one single HgTe quantum well ($S$ = 10~mV/kW for $\lambda$=148~$\mu $m) 
has about 100 times higher sensitivity per quantum well than devices based on GaAs QWs~\cite{footnote}.
The sensitivity  of HgTe QW based devices 
can be further  improved simply by using a larger number of QWs.

To determine the time resolution of our device we used  3~ps pulses of  FELIX.
The left inset in Fig.~\ref{fig2} shows that 
the response time is at most 100~ps. 
We attribute  the observed time constant  to the bandwidth  
of our electronic set-up.  A fast response is typical for photogalvanics where the signal decay time is 
determined by  the momentum relaxation time~\cite{book,Dyakonov08}
being in our samples  of the order of  $0.3$~ps at room temperature.  
As a  large dynamic range is important for detection of laser radiation, we investigated
the dependence of the sensitivity of the detection system on the radiation intensity applying
$cw$ and high-power pulsed radiation. We observed that the ellipticity detector at homogeneous irradiation 
of the whole sample remains linear up to 2~MW/cm$^{2}$ over more than nine orders of magnitude.

\begin{figure}[h]
\includegraphics[width=0.9\linewidth]{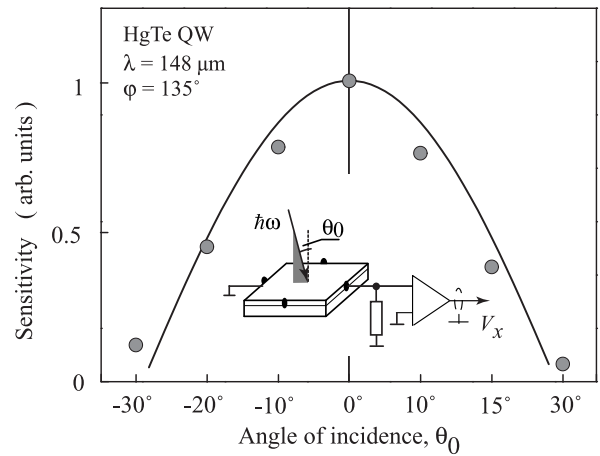}
\caption{Sensitivity of the  (013)-oriented HgTe QW CPGE-detector unit  as a 
function of the angle of incidence for circularly polarized radiation. 
The data are shown for the rotation of the angle of incidence in ($xz$)-plane.
Full line shows the fit of the data 
to the phenomenological theory taken into account besides CPGE  an additional 
contribution of the circular photon drag effect~\protect \cite{CPDE}.
}
 \label{fig4}
\end{figure}%

In a further experiment we checked the variation of the sensitivity due to a
deviation from normal incidence of  radiation.
Measurements  at oblique incidence demonstrate that  the CPGE  current attains 
maximum at normal incidence. This is  shown in Fig.~\ref{fig4}  
for angles of incidence $\theta_0$ in the range $\pm 30$ degrees at a  wavelength 
$\lambda = 148~\mu$m and for circularly polarized light. We find that 
the sensitivity is  reduced at a deviation 
from normal incidence. This behaviour follows from  Eq.~\eqref{photogalv}
yielding $j \propto t_p t_s \cos{\theta}$ where $t_p$ and $t_s$ are the Fresnel 
transmission coefficients for $s$- and $p$-polarized light, and $\theta$ is the refraction angle given by $\sin{\theta}=\sin{\theta_0}/n_\omega$, where $n_\omega$ is the refraction index~\cite{IvchenkoPikus}. Experimentally, 
however, we observed that in our samples the photocurrent drops more strongly with 
increasing $\theta_0$ than expected from Fresnel's formula. We  attribute 
this effect  to a superposition  of CPGE with  the circular photon drag 
effect~\cite{CPDE} at oblique incidence.  Because both effects are proportional 
to $P_{\rm circ}$ and have the same temporal kinetic, this superposition does 
not compromise  the detector's functionality.

\section{Detection of the ellipse's azimuth angle with bulk G\lowercase{a}A\lowercase{s}}

To obtain all Stokes parameters besides the CPGE which provides an information
on the radiation helicity, an additional detector element for determination of 
the ellipse azimuth is needed~\cite{APL2007}.
In our previous work aimed at the detection of THz radiation we used for this purpose 
the linear photogalvanic effect in SiGe  QWs.  The LPGE in SiGe 
structures is also detected in the mid-infrared range~\cite{PRB2002}. However, the signal 
is rather small and the response is obtained only at wavelength about 10~$\mu$m under resonant intersubband transitions.  
To search for other materials, with the aim to increase the sensitivity and to extend the spectral range of the detector, 
we analyze here the LPGE in bulk GaAs crystals. 
Bulk GaAs crystals belong to  T$_d$ point group symmetry. While the CPGE in materials 
of this symmetry is forbidden, an  LPGE current can be generated applying linearly or elliptically polarized radiation.

\begin{figure}[h]
\includegraphics[width=0.9\linewidth]{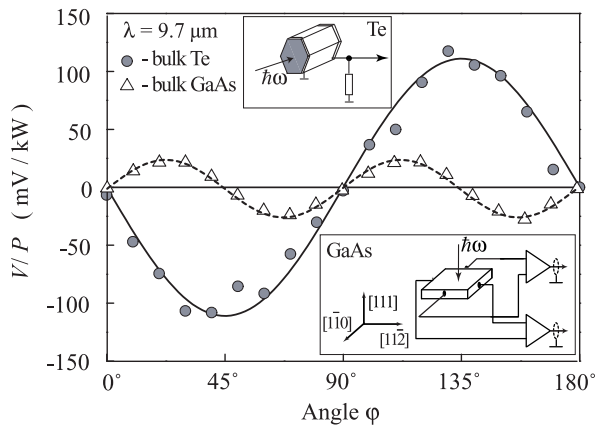}
\caption{Helicity dependence of the photoresponse normalized by the 
radiation power $V/P$ in bulk Te (circles) and bulk GaAs (triangles) samples
at room temperature. The signals are obtained 
at $\lambda = 9.7~\mu$m applying pulsed  TEA CO$_2$ laser radiation.
Full and dashed lines are fits after 
$V_z/P\propto \sin{2\varphi}$ and $V_{[1 \bar{1} 0]}/P \propto \sin{4\varphi}$ for Te and GaAs, respectively.
Insets show the experimental geometry for both samples. In the GaAs-detector unit 
the signal of each contact pair is fed into a differential amplifier
floating against ground.
}
 \label{fig5}
\end{figure}%

We prepared (111)-oriented $p$-type GaAs samples of $5\times 5 \times 2$~mm$^3$ sizes 
with a contact pair along $x \parallel [1 \bar{1} 0]$  
and $y \parallel [11\bar{2}]$ axes as sketched in Fig.~\ref{fig5}.
We used materials with free hole densities 
of about $2.3 \times 10^{16}$~cm$^{-3}$ like it previously been used for 
detection of the plane of polarization of linearly polarized
radiation~\cite{Andrianov88}. In our measurements we aligned 
the [1$\bar{1}$0]-side of the sample parallel to the light polarization 
vector of the laser beam.
The current of each contact pair is fed into a differential amplifier
floating against ground.

According to 
Eq.~(\ref{photogalv})
irradiation with polarized radiation propagating in the [111] crystallographic direction
yield transverse signals   given by
\begin{eqnarray}
\label{Td}
\frac{V_{[1 1 \bar{2}]}}{P}  &=& C(\omega) \frac{ (|E_x|^2 - |E_y|^2)}{|E|^2}, \nonumber \\
\frac{V_{[1 \bar{1} 0]}}{P} &=& C(\omega) \frac{ (E_x E_y^* + E_y E_x^*)}{|E|^2}, 
\end{eqnarray}
where $C(\omega)$ is a constant factor being proportional to the only one non-zero component 
of the third rank tensor $\bm{\chi}$~\cite{IvchenkoPikus}.
In our experimental set-up where the radiation ellipticity is varied by the rotation 
of the quarter-wave plate by an angle $\varphi$, $(|E_x|^2 - |E_y|^2) / |E|^2 =  (1+\cos{4\varphi})/2$
 and $(E_x E_y^* + E_y E_x^*)/|E|^2 =  \sin{4\varphi}/2$.
Applying infrared radiation we clearly detected both dependences of the photosignal.
Figure~\ref{fig5} shows the angle $\varphi$ dependence of $V_{[1 \bar{1} 0]}/P$  obtained 
applying radiation at $\lambda = 9.7$~$\mu$m.

From the angle $\varphi$ dependence
we obtain the LPGE constant $C(\omega)$ which determines the detector element sensitivity 
providing the calibration of the device.
Figure~\ref{fig2} shows the 
spectral behaviour of the constant $C(\omega)$
at infrared wavelengths obtained by the Q-switched CO$_2$ laser. 
The sensitivity of the order of 20~mV/kW is measured 
for the angle $\varphi = 22.5^\circ$ at which signal achieves its maximum value.
This value is comparable to that of the HgTe QW CPGE-detector unit.
We also obtained a considerable signal in the terahertz
range characterized by the same angle $\varphi$ dependences.
Here the sensitivity  at 148~$\mu$m is also comparable to that of the HgTe QW CPGE-detector unit 
and is about 20 times higher than that measured previously in SiGe QWs~\cite{JAP08}.  

To obtain the Stokes parameters of a radiation field the signals from HgTe QW and 
bulk GaAs must be measured simultaneously. In the work on a THz ellipticity detector 
we simple stacked the CPGE- and the LPGE-detector units. This arrangement is only 
possible if the first of the two elements is practically transparent to the incident radiation.  
Therefore we carried out FTIR measurements of the transmission of HgTe QW. These data are 
shown in  insets to Figs.~\ref{fig2} and~\ref{fig3}. The essential result is that 
the sample is almost transparent and can be used in a stacked configuration.
The  transmittance in the whole spectral range is
about 30-40 percents which just corresponds to the reflectivity of the sample.
The oscillations in  the  spectrum are due to  interference
in the plane-parallel transparent semiconductor slab. The magnitude of the reflection 
can be reduced  by anti-reflection coatings improving the sensitivity 
of the detector system.

\section{Stokes parameters}

From the signals obtained by the CPGE-detector unit (HgTe QW) and
by the LPGE-unit (bulk GaAs) it  follows that simultaneous
measurements of two signals allow one the 
determination 
of the Stokes parameters which completely 
characterize the state of polarization
of the radiation field.
The Stokes parameters defined according to~\cite{BornWolf} are directly 
measured in our detector 
units yielding
\begin{eqnarray}
\label{Stockes1}
s_0 &=& |E|^2, \nonumber \\
\frac{s_1}{s_0} &=& \frac{|E_x|^2 - |E_y|^2}{|E|^2} = \frac{V^{\rm GaAs}_{[1 1 \bar{2}]}}{P\cdot C(\omega)} ,\\
\frac{s_2}{s_0} &=& \frac{E_x E_y^* + E_y E_x^*}{|E|^2} = \frac{V^{\rm GaAs}_{[1\bar{1}0]}}{P\cdot C(\omega)} ,\nonumber \\
\frac{s_3}{s_0} &=& \frac{{\rm i}(E_y E_x^* - E_x E_y^*)}{|E|^2}= P_{\rm circ} = \frac{V_x^{\rm HgTe}}{P\cdot S(\omega)},\nonumber 
\end{eqnarray}
where the  parameter $s_0 \propto P$ does not contain information about the 
polarization state and determines the radiation intensity.

\section{CPGE unit based on bulk T\lowercase{e}}

Finally we investigated CPGE in another narrow band semiconductor --- bulk Te --- 
with respect to its applicability in an ellipticity detector and
compare the results with the data on  HgTe QW structures.
In fact bulk Te was the first material in which the CPGE was detected~\cite{Asnin78p74}.  
Single crystal Te belongs to the point group D$_3$ where second rank pseudotensors are diagonal 
in a coordinate system with $z$ along the trigonal crystallographic axis. This axis is also the 
growth direction of bulk crystals. The macroscopic shape of Te is hexagonal. Thus, irradiating 
a sample along $z$ direction generates a longitudinal  current along this axis. Equation~\eqref{photogalv} 
yields for the photosignal $V_z \propto j_z$ in Te \cite{Ivchenko78p74}
\begin{equation}
\label{Te}
V_z/P = \tilde{S}(\omega) P_{\rm circ} \,. 
\end{equation}
In order to measure this current we prepared two contacts around circumferences 
of hexagonal prism shaped specimens, 
see inset in Fig.~\ref{fig5}. 
We used  $p$-doped tellurium crystals which at room 
temperature have the hole concentration of about $3.5 \times 10^{16}$~cm$^{-3}$. 
The samples were of 11.5~mm long and and had a cross section of $\approx 20$~mm$^2$. 
Illuminating the crystal along $z$ axis
by mid-infrared radiation we detected a polarization dependent signal 
reversing polarity upon switching the 
radiation helicity (see Fig.~\ref{fig5}). 
In our experimental set-up the radiation helicity is varied by the rotation 
of the quarter-wave plate by an angle $\varphi$ according to $P_{\rm circ}=\sin 2\varphi$.
The signal $V_z$ is well described 
by this dependence. The possible contribution of the longitudinal 
photon drag effect is vanishingly small in our samples at room 
temperature in agreement to Ref. \onlinecite{Asnin78p74}.  Figure~\ref{fig2} 
shows the spectral behaviour of the sensitivity of the Te sample 
obtained for circularly polarized radiation. 
Our investigations demonstrate that in this material the sensitivity 
in the mid-infrared  is about three times higher than that observed 
for a HgTe single QW (Fig.~\ref{fig2}). The data prove that bulk Te 
can be used as CPGE detector unit in the mid-infrared. However,   
application of multiple HgTe QWs for which the sensitivity scales linearly with the number 
of QWs should provide a sensitivity substantially higher than that of Te making HgTe QWs preferable for ellipticity detection. 
In the THz-range we did not find a CPGE signal in Te even when applying the highest available radiation power.

\section{Conclusion}
 
We demonstrated that the application of novel narrow gap semiconductor 
low dimensional materials 
allows all-electric detection of the Stokes parameters of elliptically 
polarized radiation from the mid-infrared to the THz range. The sensitivity 
and linearity of the detection system developed here has been found 
to be sufficient to characterize the polarization of laser radiation from low-power cw-lasers to high-power 
laser pulses. The short relaxation time of free carriers   in semiconductors at room 
temperature makes it possible to detect sub-ns laser pulses demonstrated with the  
free-electron-laser FELIX.
Finally we note that a further increase of the sensitivity may be obtained by using 
multiple QWs and technologically available (112)-grown samples.

\acknowledgments We thank S.A. Tarasenko, E.L. Ivchenko and V.V.~Bel'kov for useful discussions.
The financial support of the DFG and RFBR is gratefully acknowledged.
Work of L.E.G. is also supported by ``Dynasty'' Foundation --- ICFPM and President 
grant for young scientists.

\newpage


\begin{thebibliography}{99}
\bibitem{footnote}  We note that in~Ref.\protect \onlinecite{JAP08} applying GaAs structures made of thirty   QWs
we observed $\approx$~3~mV/kW at $\lambda$=148~$\mu$m corresponding to 0.1~mV/kW for a single QW.
\end{thebibliography}
\end{document}